\documentclass[prb]{revtex4-1}%
\usepackage{amssymb}
\usepackage{amsmath}
\usepackage{graphicx}
\pdfoutput=1
\usepackage{amsfonts}%
\begin{document}
\title{Dispersive magnetometry with a quantum limited SQUID parametric amplifier}
\author{M. Hatridge$^{1}$, R. Vijay$^{2}$, D. H. Slichter$^{2}$, John Clarke$^{1}$,
and I. Siddiqi$^{2}$}
\affiliation{$^{1}$Department of Physics, University of California, Berkeley, CA 94720 and
Materials Sciences Division, Lawrence Berkeley National Laboratory, Berkeley,
CA 94720}
\affiliation{$^{2}$Quantum Nanoelectronics Laboratory, Department of Physics, University of
California, Berkeley, CA 94720}

\begin{abstract}
%\large
There is currently fundamental and technological interest in measuring and
manipulating nanoscale magnets, particularly in the quantum coherent regime.
To observe the dynamics of such systems one requires a magnetometer with not only
exceptional sensitivity but also high gain, wide bandwidth and low backaction.
We demonstrate a dispersive magnetometer consisting of a two-junction SQUID in
parallel with an integrated, lumped-element capacitor. Input flux signals are
encoded as a phase modulation of the microwave drive tone applied to the
magnetometer, resulting in a single quadrature voltage signal. For strong
drive power, the nonlinearity of the resonator results in quantum limited,
phase sensitive parametric amplification of this signal, which improves flux
sensitvity at the expense of bandwidth. \ Depending on the drive parameters,
the device performance ranges from an effective flux noise of 0.29\textbf{
}$\mu\Phi_{0}$Hz$^{-\frac{1}{2}}$ and 20 MHz of signal bandwidth to a noise of
0.14 $\mu\Phi_{0}$Hz$^{-\frac{1}{2}}$ and a bandwidth of 0.6 MHz. These
results are in excellent agreement with our theoretical model.

\end{abstract}
\maketitle
%\large
\section{Introduction}

The dc Superconducting Quantum Interference Device (SQUID)---a superconducting
loop interrupted by two Josephson junctions---is an extremely sensitive
detector of magnetic flux, and has been used in a wide variety of applications
for almost half a century \cite{Clarke2004,Clarke2006}. Recent progress in
nanoscale magnets \cite{Bader2006} has generated excitement about using
magnetic molecules for both classical and quantum information storage and
processing
\cite{Leuenberger2001,Tejada2001,Gallop2002,Lam2003,Gatteschi2006,Cleuziou2006,
Troeman2007, Granata2008, Wernsdorfer2009}. In principle, a SQUID has
sufficiently high intrinsic flux sensitivity and bandwidth to resolve the spin
state of a single magnetic molecule. However, a conventional dc SQUID with
resistively shunted Josephson tunnel junctions is operated with a current bias
just above its critical current, and the continuous dissipation in the shunt
resistors produces local heating and a backaction that can potentially induce
relaxation and decoherence in a nanoscale magnet. This dissipation can be
eliminated by using an unshunted SQUID and applying fast current pulses to
measure its critical current \cite{vanderWal2000,Wernsdorfer2009}. However,
the flux sensitivity is significantly lower than in the resistively shunted
case \cite{vanharlingen1982} because the repetition rate---and hence the
bandwidth---are limited by the time ($\sim$1 ms for aluminum tunnel junctions at millikelvin temperatures) required for the SQUID to
cool to its equilibrium temperature after returning to the zero-voltage state
\cite{Segall2004}. Any backaction associated with switching to the voltage
state remains.

Alternatively, the SQUID can be operated in the superconducting regime where
it functions as a flux dependent nonlinear inductor, and forms a nonlinear
resonator when shunted with a capacitor [Fig. 1(a)]. In our device, we apply a
fixed frequency microwave drive to this resonator and demodulate the reflected
microwave signal. An input flux signal results in a variation of the resonance
frequency and a corresponding phase modulation of the microwave drive tone. At
specific bias points in the presence of a sufficiently intense drive tone,
parametric amplification occurs and the flux sensitivity is enhanced.
Dispersive SQUID techniques have been studied in a variety of different
microwave circuit configurations over the past thirty years
\cite{Likharev1985,Kuzmin1985,mates2008}. Recent work on the dispersive
readout of superconducting qubits---single, pseudospin-1/2 systems---also
harnesses the nonlinearity of the Josephson junction to boost sensitivity, but
typically these devices are operated in the bistable regime as digital
detectors \cite{PhysRevLett.96.127003,Siddiqi2006,Mallet2009}. In this
article, we demonstrate an analog magnetometer with megahertz bandwidth
suitable for measuring transitions between states in multilevel spin systems
\cite{delBarco2004} and the macroscopic magnetization of spin ensembles
\cite{Jamet2000}. Depending on the operating conditions, the performance
ranges from an effective flux noise of 0.14\textbf{ }$\mu\Phi_{0}$%
Hz$^{-\frac{1}{2}}$ and 0.6 MHz of signal bandwidth to a noise of 0.29
$\mu\Phi_{0}$Hz$^{-\frac{1}{2}}$ and a bandwidth of 20 MHz.  
This performance results from the large gain, bandwidth and nearly quantum limited noise temperature 
of the parametric amplifier which by itself is suitable for a variety of dispersive measurements
as a general purpose amplifier. These results are
in quantitative agreement with our theoretical model which, in particular,
predicts that low flux noise and wide bandwidth are obtained for a low Q (quality factor)
resonator. Our theory allows us to optimize our device for
specific applications, and provides insight into the fundamental and practical
limitations of a single SQUID operated in the dispersive regime.

\section{Theory}

We model our magnetometer as consisting of two stages: a transducer which
upconverts a low frequency magnetic flux signal to a microwave voltage signal
and a subsequent parametric gain stage [see Fig. 1(b)]. Using this picture, we
derive an expression for the flux sensitivity based on the circuit parameters
and the parametric gain. \ We first consider the dynamics of the Josephson
oscillator. \ The supercurrent $I(t)$ flowing through a Josephson tunnel junction is
related to the phase difference $\delta\left(  t\right)  $ across it by
$I(t)=I_{0}\sin\delta(t)$, where $I_{0}$ is the critical current. \ For a
SQUID with loop inductance $L\ll\Phi_{0}/2I_{0}$, the\ critical current is $I_{c}\left(  \Phi\right)  =2I_{0}\left\vert \cos(\frac{\pi\Phi}%
{\Phi_{0}})\right\vert $, where $\Phi$ is the flux through the SQUID\ loop and
$\Phi_{0}\equiv h/2e$ is the magnetic flux quantum.\ \ Thus, we treat the
SQUID\ as a junction with a flux dependent critical current. In our experiment
the SQUID\ is shunted with a lumped-element capacitor, forming an electrical
resonator with resonant frequency $\omega_{p0}\left(  \Phi\right)  /2\pi
=\sqrt{2\pi I_{c}\left(  \Phi\right)  /(\Phi_{0}C)}/2\pi$. \ The resonator is
connected directly to a microwave transmission line of characteristic
impedance $Z_{0}$ [Fig. 1(a)], resulting in a quality factor $Q=\omega_{p0}%
Z_{0}C$.

The dynamics of this system are well described by the Duffing equation, in
which the sinusoidal current phase relationship of the junction is truncated
after the first nonlinear term \cite{Dykman1980}:%

\begin{equation}
\frac{\partial^{2}\delta}{\partial t^{2}}+2\Gamma\frac{\partial\delta
}{\partial t}+\omega_{p0}^{2}(\Phi)(\delta-\frac{\delta^{3}}{6})=\frac{2\pi
}{\Phi_{0}C}I_{d}\cos(\omega_{d}t).\label{eqn_duffing}%
\end{equation}
Here, $\Gamma=\left(  2Z_{0}C\right)  ^{-1}$ and $I_{d}$ is the amplitude of
the rf drive at frequency $\omega_{d}/2\pi$. Next we consider a flux
$\Phi=\Phi_{b}+\Delta\Phi\cos(\omega_{s}t)$ applied to the SQUID, where
$\Phi_{b}$ is a static flux bias and $\Delta\Phi\ll$ $\Phi_{0}$ is the
amplitude of a weak flux signal at frequency $\omega_{s}$.\ We calculate the
system response by assuming a solution of the form $\delta(t)=\delta_{0}%
\cos(\omega_{d}t-\theta)+\epsilon(t)$, where the first term is the steady
state solution for $\Delta\Phi=0$ and $\epsilon(t)$ is a small perturbation of
the junction phase due to $\Delta\Phi$, and substituting it into Eq.
(\ref{eqn_duffing}). The resulting expression for the junction phase
perturbation $\epsilon(t)$ is%

\begin{align}
&  \frac{\partial^{2}\epsilon}{dt^{2}}+2\Gamma\frac{\partial\epsilon}%
{dt}+\epsilon\omega_{p0}^{2}\left(  1-\frac{\delta_{0}^{2}}{4}\right)  \left[
1-\frac{\delta_{0}^{2}}{4-\delta_{0}^{2}}\cos\left(  2\omega_{d}%
t-2\theta\right)  \right] \label{eqn_paramp}\\
&  =\frac{\Delta\Phi}{Z_{0}C\Phi_{0}}\frac{\partial V_{rf}}{\partial\Phi
}\left[  \cos(\omega_{d}t+\omega_{s}t-\theta)+\cos(\omega_{d}t-\omega
_{s}t-\theta)\right]  .\nonumber
\end{align}
We recognize the left hand side of Eq. (\ref{eqn_paramp}) as the equation for
a parametrically driven harmonic oscillator. For appropriate bias conditions,
the system can amplify \cite{Feldman1975,Yurke1989,Vijay2008a} any additional weak rf\ signal with frequency
$\omega_{rf}/2\pi$ near $\omega_{d}/2\pi$. \ In this doubly degenerate mode of
parametric amplification \cite{Vijay2008a}, a single sideband signal at frequency $\omega
_{rf}/2\pi$ is amplified with a voltage gain $\sqrt{G}$, and an
idler signal is also produced at frequency $\omega_{i}/2\pi=(2\omega
_{d}-\omega_{rf})/2\pi$ with gain $\sqrt{G-1}$ and phase factor $e^{i\phi}$
[Fig. 2(a)]. In the high gain limit, the voltage signal-to-noise ratio (SNR) is
degraded by a factor of at least $\sqrt{2}$, since the amplified signal is
accompanied by incoherent noise from both the signal and idler frequencies.
\ For an operating temperature $T\ll T_{Q}=\hbar\omega_{d}/2k_{B}$, this noise
is set by the amplitude of quantum fluctuations at frequency $\omega_{d}$ and
the amplifier is quantum limited with a noise temperature $T_{N}=T_{Q}$. Other
Josephson junction based parametric amplifiers have been shown to operate with
near quantum limited noise temperature \cite{Castellanos-Beltran2008,
Bergeal2009}. \ If such an amplifier is now presented with a double sideband
signal, symmetric about the drive tone with coherent components at both the
signal and idler frequencies, the output voltage is a coherent
combination of these two signals. \ We can express this double sideband signal
in terms of two orthogonal quadrature signals---one of which is amplified
and the other deamplified. \ This process is shown schematically for the
amplified quadrature in Fig. 2(b). \ If the signal lies fully along the
amplified quadrature ($\alpha=0$), it is amplified without adding additional noise, a
process known as phase sensitive amplification \cite{Caves1982}. \ 

Examining the right hand side of Eq.(\ref{eqn_paramp}), we see that the flux
signal at frequency $\omega_{s}/2\pi$ has been parametrically upconverted
through interaction with the drive tone, resulting in a double sideband rf
signal with components at frequencies $\left(  \omega_{d}\pm\omega_{s}\right)
/2\pi$ which can be expresssed as a single quadrature signal with angle
$\alpha=\theta$ relative to the drive tone. \ The amplitude of these two signals can be
computed using the transduction coefficient%

\[
\frac{\partial V_{rf}}{\partial\Phi}=\pi\frac{2I_{0}Z_{0}}{4\Phi_{0}}%
\sin\left(  \frac{\pi\Phi_{b}}{\Phi_{0}}\right)  \left(  \delta_{0}%
-\frac{\delta_{0}^{3}}{8}\right)  .\ \ \ \ \ \ \ \ \ \left(  -\frac{\Phi_{0}%
}{2}<\Phi_{b}<\frac{\Phi_{0}}{2}\right)  \
\]

In the limit of low drive current $I_{d}$, $\sqrt{G}=1$ and the paired rf
signals serve to modulate the phase of the reflected microwave drive signal at
frequency $\omega_{s}/2\pi$. \ In the high gain limit, the component of the
transduced signal which lies along the amplified quadrature of the phase
sensitive amplifier is noiselessly amplified with gain $2\sqrt{G}$. \ The
transduced signal, however, does not lie fully along the amplified quadrature
at high gain points, so that the effective transduction coefficient is
reduced. \ Consequently, in the high gain limit, the amplitude of the output
signal is given by $\Delta\Phi\eta\frac{\partial V_{rf}}{\partial\Phi}\sqrt
{G}$, where $\eta=2\cos\theta$. \ 

To characterize the flux sensitivity of the magnetometer, we first note that the output of our device
is further amplified by a cryogenic high electron mobility transistor (HEMT) amplifier and a room temperature amplifier before being mixed down
and digitized for further processing. The noise of this amplification chain is given by the system noise temperature $T_{sys}$, and
 includes cable losses between our device and the HEMT. We now refer the noise
of the output rf voltage signal to an effective input flux noise via the
transduction coefficient $\eta$ $\frac{\partial V_{rf}}{\partial\Phi}$. This
effective flux noise has a single-sided spectral density
\begin{equation}
S_{\Phi eff}^{1/2}(f)=\frac{\sqrt{2(k_{B}T_{sys}+\hbar
\omega_{d}/2)Z_{0}}}{\eta\frac{\partial V_{rf}}{\partial\Phi}\Phi_{0}}%
=\frac{4}{\pi}\frac{\sqrt{2(k_{B}T_{sys}+\hbar\omega_{d}/2)Z_{0}}}{2I_{0}%
Z_{0}\eta\sin\left(  \frac{\pi\Phi_{b}}{\Phi_{0}}\right)  \left(  \delta
_{0}-\frac{\delta_{0}^{3}}{8}\right)  }   \;\;\;  \frac{\Phi_{0}}{\sqrt{Hz}}.
\end{equation} \ 

Note that in the above expression $T_{sys}$ refers to the added noise of the amplification chain
and the total noise includes a contribution from zero point fluctuations given by $\hbar
\omega_{d}/2$. For weak drive amplitudes, where the resonator response is linear and $G=1$,
$T_{sys}$ is dominated by $T_{HEMT}$, the noise temperature of the HEMT amplifier and is
typically much larger than the quantum limit. \ Furthermore, we can
express the junction oscillation amplitude in terms of the voltage drive as
$\delta_{0}-\delta_{0}^{3}/{8}$ $\approx\delta_{0}\approx V_{d}%
Q/{Z_{0}2I_{0}}$. \ In the absence of parametric amplification, there is no preferred quadrature 
for amplification and we set $\eta=1$ \ to yield%

\begin{equation}
S_{\Phi eff}^{1/2}(f)\approx\frac{4}{\pi}\frac{\sqrt
{2k_{B}T_{HEMT}Z_{0}}}{\sin\left(  \frac{\pi\Phi_{b}}{\Phi_{0}}\right)
V_{d}Q}  \;\;\;  \frac{\Phi_{0}}{\sqrt{Hz}}  \text{ for }G=1, \label{eqn_fluxnoise_lin}%
\end{equation}
which varies inversely with resonator $Q$ and drive voltage $V_{d}$. \ Thus,
in this regime higher sensitivity can be achieved by increasing the $Q$ of the
resonator and the drive amplitude. \ In practice, however, the maximum drive
voltage is limited by the onset of nonlinearity inherent in any SQUID\ based
resonator \cite{Manucharyan:2007}. \ 

For strong drive amplitudes, where the resonator response is nonlinear and
$G\gg1$ such that $T_{HEMT}/G \ll \hbar
\omega_{d}/2 $, the noise temperature of the system is determined by the amplitude of
quantum fluctuations at the drive frequency,  and $T_{sys}\approx \hbar \omega_{d}/2k_{B}$.
 Further, as $G\rightarrow\infty$, $\delta_{0}$ approaches a
critical value\cite{Manucharyan:2007} of $\delta_{c}={4}/{3^{1/4}\sqrt{Q}}$%
. \ For $Q\gtrsim10$, $\delta_{c}$ $\leq1$ and we can
make the approximation $\delta_{0}-{\delta_{0}^{3}}/{8}\approx\delta_{c}$.
\ Similarly, at this operating point, the angle $\theta$ also approaches a critical value of
60$^{\circ}$, so that we can approximate $\eta\approx1$. Since the maximum value of $\eta$ is 2, only half of the transduced signal is noiselessly amplified resulting in an effective flux noise two times higher than what could be achieved in the ideal case with $\theta=0$. A similar analysis which treats the system as a degenerate parametric amplifier with a detuned pump and takes into account the non-orthogonality of the amplified and deamplified quadratures, gives the same result in this limiting case \cite{laflamme2010}. The effective flux noise computes to
\begin{equation}
S_{\Phi eff}^{1/2}(f)\approx\frac{(2\sqrt{3})^{1/2}}{\pi
}\frac{\sqrt{\hbar\omega_{d}}}{2I_{0}\sin\left(  \pi\Phi_{b}/ \Phi_{0}%
\right)  }\sqrt{\frac{Q}{Z_{0}}} \;\;\;  \frac{\Phi_{0}}{\sqrt{Hz}}   \text{ for }G\gg1. \label{eqn_fluxnoise_nl}%
\end{equation}

This expression has the remarkable feature that the effective flux noise
varies directly with the resonator $Q$, so that a resonator with lower $Q$ has
improved flux sensitivity. \ This results from the requirements of parametric
amplification, in particular that high parametric gain occurs at a critical
phase oscillation $\delta_{c}\propto Q^{-1/2}$, which in turn limits the
maximum achievable transduction coefficient. \ Additionally, lowering the
resonator $Q$ leads to parametric amplification with increased bandwidth for a
given parametric gain, and so is doubly desirable. \ In both the linear and
nonlinear regime, it is advantageous to increase the SQUID critical current
and operate at a flux bias near $\Phi_{0}/2.$ However, for sufficiently large critical current
such that $\Phi_{0}/I_{0} \sim L$, where $L$ is the inductance of the SQUID loop, the simple
expression for $I_{c}\left(  \Phi\right)$ is no longer valid and one has to use numerical simulations
to determine the change in SQUID inductance as a function of applied flux, which is often a multivalued
function. Furthermore, unlike the critical current of a dc SQUID, the inductance modulation 
with flux of a dispersive SQUID is not limited by the loop inductance provided one remains
 in one of the branches of the multivalued function.

\bigskip

\section{Experiment}

\bigskip

Our lumped element resonator is shown in a false color scanning electron
microscope image in Fig. 3(a). \ The device consists of three layers: a 250-nm
thick Nb underlayer, a 180-nm thick SiN$_{x}$ insulating layer, and an
aluminum upper layer. \ The capacitor was fabricated in a split geometry, with
both electrodes on the top Al layer connected through the Nb underlayer. The split
geometry simplifies fabrication by avoiding the use of vias. \ The
SQUID was fabricated with double-angle evaporated Al-AlO$_{x}$-Al junctions,
with $2I_{0}=4.3$ $\mu$A. \ A short-circuit terminated coplanar waveguide
transmission line was also fabricated on chip to allow us to apply oscillating
flux signals\ to the SQUID\ loop. \ Static flux was provided by a
superconducting wire-wound coil. \ The device was cooled in superconducting
and Cryoperm shields, and its performance measured at 30 mK in a cryogen-free
dilution refrigerator. \ \ All static and rf lines were heavily attenuated and
filtered to minimize external noise. \ 

To determine the dependence of the resonant frequency on applied flux, we used a
vector network analyzer to measure the phase of a weak microwave tone
reflected from the resonator as a function of frequency and flux. \ The
results are shown in Fig. 3(b). \ As the flux through the SQUID\ washer was
varied, the resonant frequency varied from a maximum of 7.2 GHz to a
minimum of 4 GHz, set by the low frequency cutoff of the circulators used in
the measurement chain. \ The horizontal bands apparent in the plot are due to
the finite directivity of the circulator used to separate the incoming and
outgoing microwave signals from the resonator. \ The sample was flux biased at
$\Phi_{b}=0.3\Phi_{0}$, where the flux sensitivity was high and the frequency
band was clear of ripples which could obscure the response to rf and magnetic
signals. \ 

We next examined the performance of the parametric amplification stage. \ To
characterize the gain, a strong rf drive tone at frequency $\omega_{d}/2\pi$
was applied to the resonator simultaneously with a weak rf signal at frequency
$\omega_{rf}/2\pi$ [Fig. 2(a)]. \ The reflected rf signal was further
amplified, demodulated by a double sideband mixer with local oscillator
frequency $\omega_{LO}/2\pi=\omega_{rf}/2\pi+110$ Hz, and digitized. \ The
gain was determined by calculating the ratio of the reflected microwave signal
with the drive tone turned on versus a calibration sweep with the drive tone turned off;
in the latter mode the signal tone is reflected from the resonator with unity gain.
\ The gain is plotted in Fig. 4(a) as a function of the frequency offset
$(\omega_{rf}-\omega_{d})/2\pi$ for different drive powers. \ The advantage of
low resonator Q is evident in these data where a gain of 15 dB is demonstrated
with a -3 dB full-bandwidth of 40 MHz. We note that the observed gain and
bandwidth deviate slightly from the analytical solution of the simple Duffing
oscillator prediction for a given drive amplitude. This discrepancy is
resolved using numerical simulations which indicate gain suppression due to
noise rounding in nonlinear resonators with very low Q.

In separate measurements using a hot/cold load, we found the system noise
temperature at the plane of the resonator to be between 29 and 37.5 $%
%TCIMACRO{\unit{K}}%
%BeginExpansion
\operatorname{K}%
%EndExpansion
$, with the scatter arising from the uncertainty in the attenuation of the hot
load line. By measuring the SNR\ improvement of rf signals measured at finite
gain versus $G=0$ dB, we calculate the system noise temperature as a function
of drive power [shown in Fig. 4(b)]. \ The lowest system noise temperature was
measured with $G=32$ dB at $\omega_{d}/2\pi=5.56%
%TCIMACRO{\unit{GHz}}%
%BeginExpansion
\operatorname{GHz}%
%EndExpansion
$, with $T_{sys}$ between $0.14$ and $0.21%
%TCIMACRO{\unit{K}}%
%BeginExpansion
\operatorname{K}%
%EndExpansion
$, corresponding to a nearly quantum limited added noise of 0.50 to 0.78
photons. At the highest gain point, the noise temperature was degraded due to
instabilities associated with operation near the critical point
\cite{Bryant1991}. \ 

We subsequently investigated the flux response of the system by applying a
flux tone at frequency $\omega_{s}/2\pi$, chosen so that the upper sideband of
the upconverted rf output at frequency $\left(  \omega_{d}+\omega_{s}\right)
/2\pi$ was offset by 10 Hz from an additional applied rf signal at frequency
$\omega_{rf}/2\pi$. This allows simultaneous measurements of the parametric
gain and flux sensitivity. \ The output signals were again amplified,
demodulated with a double sideband mixer with local oscillator frequency
$\omega_{LO}/2\pi=\omega_{rf}/2\pi+110$ Hz, and digitized. \ The SNR of the
flux response at each bias point was calculated by comparing the height of the
transduced output signal to the average of the white noise in a bandwidth of
200 Hz. \ This SNR was converted into an effective flux noise using the known
amplitude of the flux signal applied to the magnetometer. \ In these
measurements, the primary source of uncertainty in the effective flux noise is
the calibration of the flux signal, which we estimate to be a few percent.
\ In Fig. 4(c), we plot the effective flux noise as a function of drive
amplitude and flux signal frequency. \ In the linear regime, the magnetometer
bandwidth is limited by the bandwidth of our flux excitation line and is demonstrated to be
at least 80 MHz. Using our value of $Q\approx26$, we expect the bandwidth
to be greater than 100 MHz.  By biasing the
resonator in the nonlinear regime, we trade bandwidth for parametric gain and
reduced noise. By operating at a parametric gain of 32 dB, we achieved a
minimum effective flux noise of 0.21 $\mu\Phi_{0}$Hz$^{-\frac{1}{2}}$ at
$\omega_{s}/2\pi=100$ kHz. \ However, this is not the lowest attainable noise
as the demodulated noise in these measurements with $\omega_{LO}\neq\omega
_{d}$\ is the sum of incoherent noise sidebands above and below the LO
frequency, thus degrading the effective flux noise.

To determine the optimum device performance, we performed a second set of
measurements with $\omega_{LO}=$ $\omega_{d}$. \ With this demodulation
technique we made use of the single quadrature nature of the transduced flux
signal. The effective flux noise as function of drive power measured at
$\omega_{s}/2\pi=$100 $%
%TCIMACRO{\unit{kHz}}%
%BeginExpansion
\operatorname{kHz}%
%EndExpansion
$ for both demodulation techniques is shown in Fig. 4(c) and clearly
demonstrates the advantage of demodulating with $\omega_{LO}=$ $\omega_{d}$.
In the linear regime at low powers, the latter\ demodulation scheme improves the 
effective flux noise by a factor of 2 since both sidebands containing information are used.
\ At high drive powers the improvement is only $\sqrt{2}$, since both sidebands contain identical
signal and noise due to parametric amplification. \ The minimum
effective flux noise of 0.14 $\mu\Phi_{0}$Hz$^{-\frac{1}{2}}$ was achieved
with $G=32$ dB, with a flux signal bandwidth of 600 kHz set by the parametric
amplifier half bandwidth. If we substitute our system parameters into  Eq.(\ref{eqn_fluxnoise_nl}), we predict a minimum effective flux noise of 0.14
$\pm$ $0.007$ $\mu\Phi_{0}$Hz$^{-\frac{1}{2}}$, in very good quantitative
agreement with our measured results. \ The uncertainty in this calculation is
the estimated uncertainty in our knowledge of the circuit parameters. \ As a
final note, we observe that, by reducing the parametric gain to 15 dB, we
achieved an effective flux noise of 0.29 $\mu\Phi_{0}$Hz$^{-\frac{1}{2}}$
while signficantly increasing the flux signal bandwidth to 20 MHz. \ 

\bigskip

\section{Conclusion}

We have demonstrated a dispersive magnetometer based on a lumped-element,
nonlinear resonator involving a two-junction SQUID. The bandwidth and
sensitivity of the device can be dynamically altered to suit the needs of specific
measurements by changing the microwave drive. \ We have achieved an 
effective flux noise of 0.14 $\mu\Phi_{0}$Hz$^{-\frac{1}{2}}$ with a bandwidth of
600 kHz. The bandwidth can be increased to 20 MHz with only a factor of 
two increase in flux noise. We expect that our magnetometer will exhibit low backaction, 
since the SQUID never switches to the voltage state, making it attractive for quantum state
measurement. There are several avenues for further improvement. The
magnetometer could be realized as two physically separate devices, a
transducer and a gain stage. This would allow for independent optimization of
the transduction coefficient and the performance of the parametric amplifier.
By increasing the Q of the transduction stage to match the bandwidth of the
parametric amplifier, and rotating the transduced signal fully into the
amplified quadrature, reductions by a factor of about 10 in the effective flux
noise should be possible. A lower effective flux noise could also be achieved
by optimizing the transduction coefficient using junctions with higher
critical currents, and by operating at a flux bias closer to $\Phi_{0}$/2. An
increased bandwidth could be obtained by reducing the noise temperature of the
microwave postamplifier, thus reducing the parametric gain necessary for
quantum noise limited operation. Moreover, the quantum noise limited
amplification we have observed suggests that this device can be used as a
general purpose first-stage rf amplifier for a variety of applications.

An attractive possibility is to replace the tunnel junctions in the SQUID with
nanobridges, provided they have sufficient nonlinearity to exhibit efficient
flux tranduction and high parametric gain\cite{vijay2009nanobridge,
vijay2010nanobridge}. Nanobridge SQUIDs have the advantage of efficient
coupling to nanoscale spin systems, such as single molecule
magnets\cite{Wernsdorfer2009}, and resilience to applied longitudinal magnetic field.

\bigskip

\section{Acknowledgements}

The authors thank O. Naaman for helpful discussions. \ This work
was supported by the Director, Office of Science, Office of Basic Energy
Sciences, Materials Sciences and Engineering Division, of the U.S. Department
of Energy under Contract No. DE-AC02-05CH11231(M.H., J.C.). \ Financial
support was also  provided by the Office of Naval Research under grant
N00014-07-1-0774 (I.S.) and AFOSR Grant FA9550-08-1-0104 (R.V., I.S.).
\ D.H.S. gratefully acknowledges support from a Hertz Foundation Fellowship
endowed by Big George Ventures.

\clearpage

\begin{figure}
[t]
\begin{center}
\includegraphics[
natheight=2.000300in,
natwidth=6.996300in,
height=2.0384in,
width=7.0629in
]%
{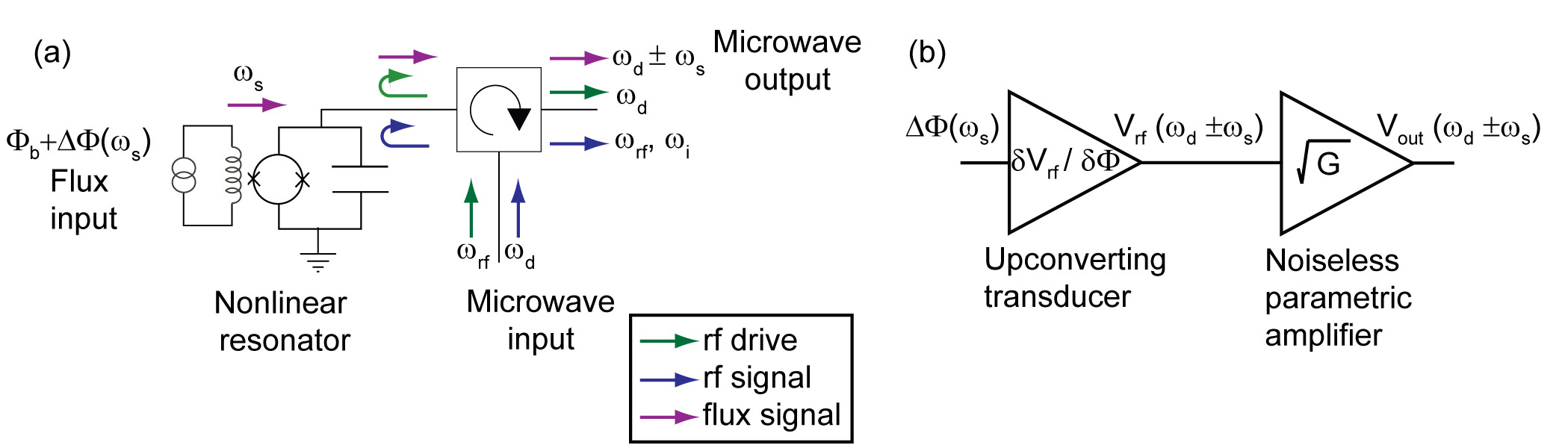}%
\caption{(a) The magnetometer consists of the nonlinear inductance of an unshunted
two-junction SQUID in parallel with an on-chip, lumped-element capacitor.
 \ An applied magnetic flux (purple arrows)\ modulates the resonant frequency, 
and is read out as a change in the phase of a microwave drive signal (green arrows) reflected from the
device through a circulator. \ If the resonator is driven strongly, the upconverted flux signal 
(purple arrows) and any additional weak rf input signal (blue arrows) will be parametrically amplified. (b) 
The magnetometer can be characterized as a dual stage
device, the first stage being an upconverting transducer of flux to microwave
voltage and the second an rf parametric amplifier.}%
\label{figure1}%
\end{center}
\end{figure}

\begin{figure}
[ptb]
\begin{center}
\includegraphics[
natheight=5.500200in,
natwidth=6.976400in,
height=5.5279in,
width=7.0041in
]%
{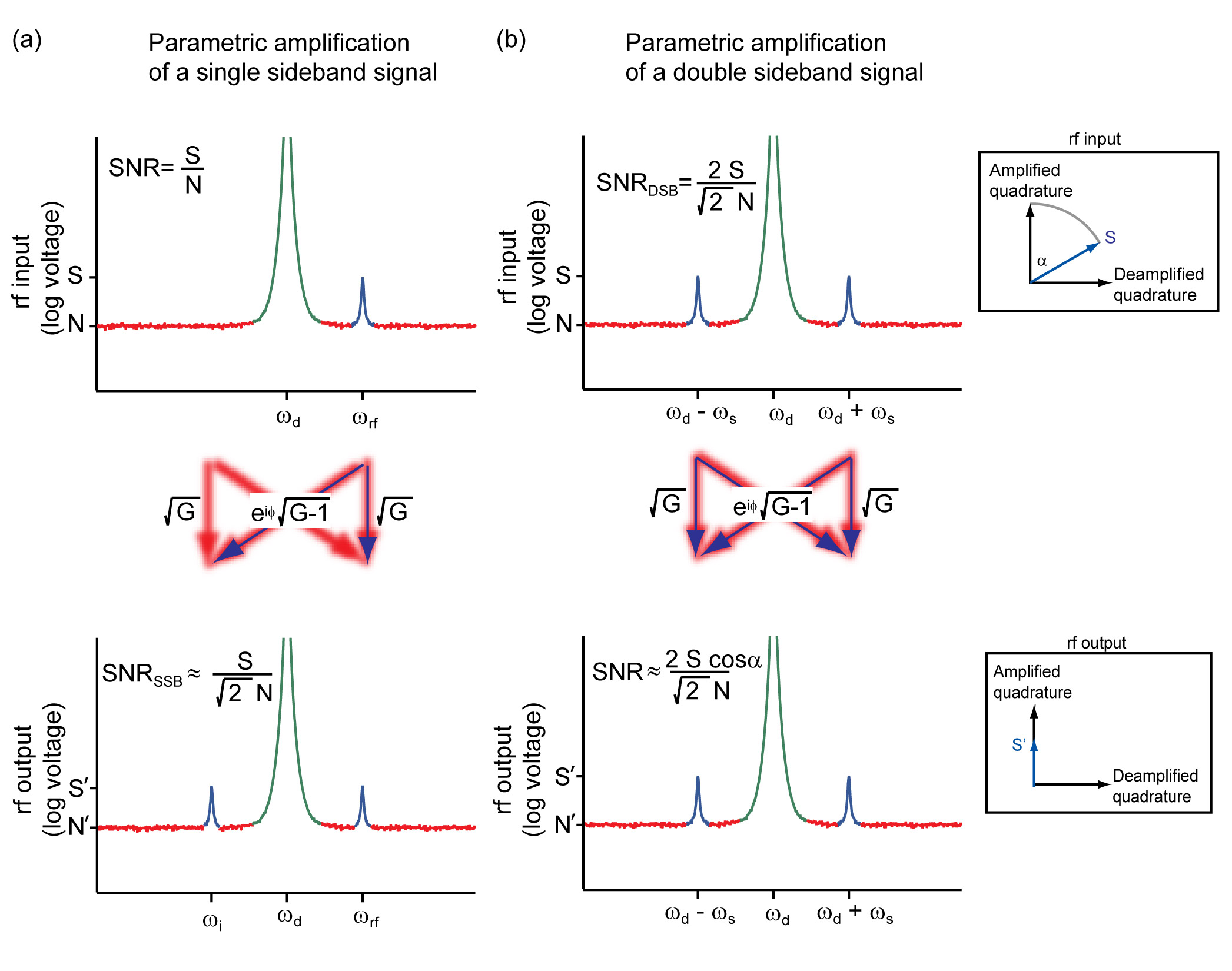}%
\caption{
Parametric amplification. The drive tone is in green, coherent
signals are in blue, and incoherent noise is in red.\ (a) A single sideband signal
at frequency $\omega_{r}f/2\pi$ is amplified with a voltage gain $\sqrt
{G}$, and an idler signal is also produced at frequency $\omega_{i}%
/2\pi=(2\omega_{d}-\omega_{rf})/2\pi$ with gain $\sqrt{G-1}$ and relative
phase $\phi$. In the high gain limit, the voltage signal-to-noise ratio
(SNR) is degraded by a factor of at least $\sqrt{2}$, since the amplfied signal
is accompanied by incoherent noise from both the signal and idler frequencies.
(b) A coherent double sideband input signal symmetric about the drive tone
results in output voltages which are a coherent combination of the signal and
idler tones. \ This process can be decomposed as amplification of two
orthogonal quadratures---one which is amplified and the other which is
deamplified. If the input signal lies fully along the amplified quadrature
($\alpha=0)$, it will be amplified without reduction of SNR.}%
\end{center}
\end{figure}

\begin{figure}
[ptb]
\begin{center}
\includegraphics[
natheight=2.50in,
natwidth=6.986800in,
height=2.5in,
width=7.0145in
]%
{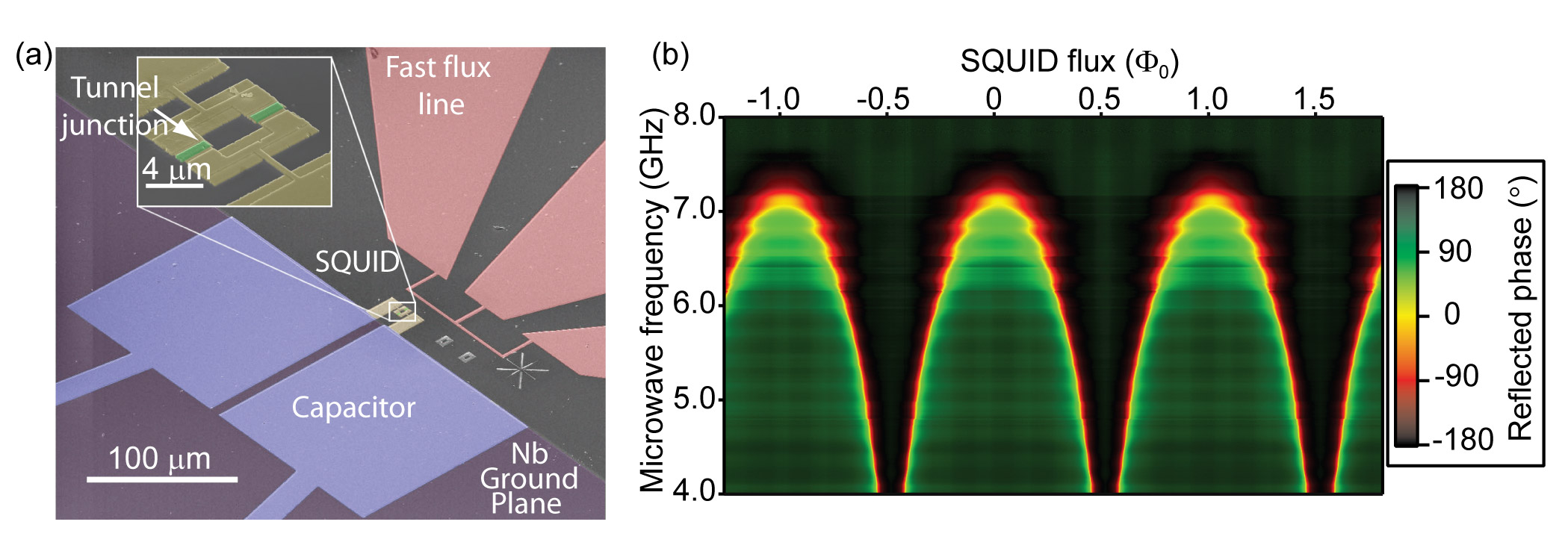}%
\caption{
(a) \ False color scanning
electron microscope\ image of the dispersive magnetometer. The SQUID\ is shown
magnified in the inset, with the Al-AlO$_{x}$-Al tunnel junctions marked in
green. \ The capacitor is formed in a split geometry, with two top layer
electrodes connected through a common niobium plane. A high bandwidth flux
line is formed by a short circuit terminated coplanar waveguide transmission
line, shown in the upper right of the figure. (b) The phase of the reflected microwave drive
is plotted versus applied SQUID\ flux and drive frequency. Each vertical slice is a
resonance curve,  with the color yellow (zero phase)  representing the resonant frequency.}%
\end{center}
\end{figure}

\begin{figure}
[ptb]
\begin{center}
\includegraphics[
natheight=5.000400in,
natwidth=6.999800in,
height=5.028in,
width=7.0275in
]
{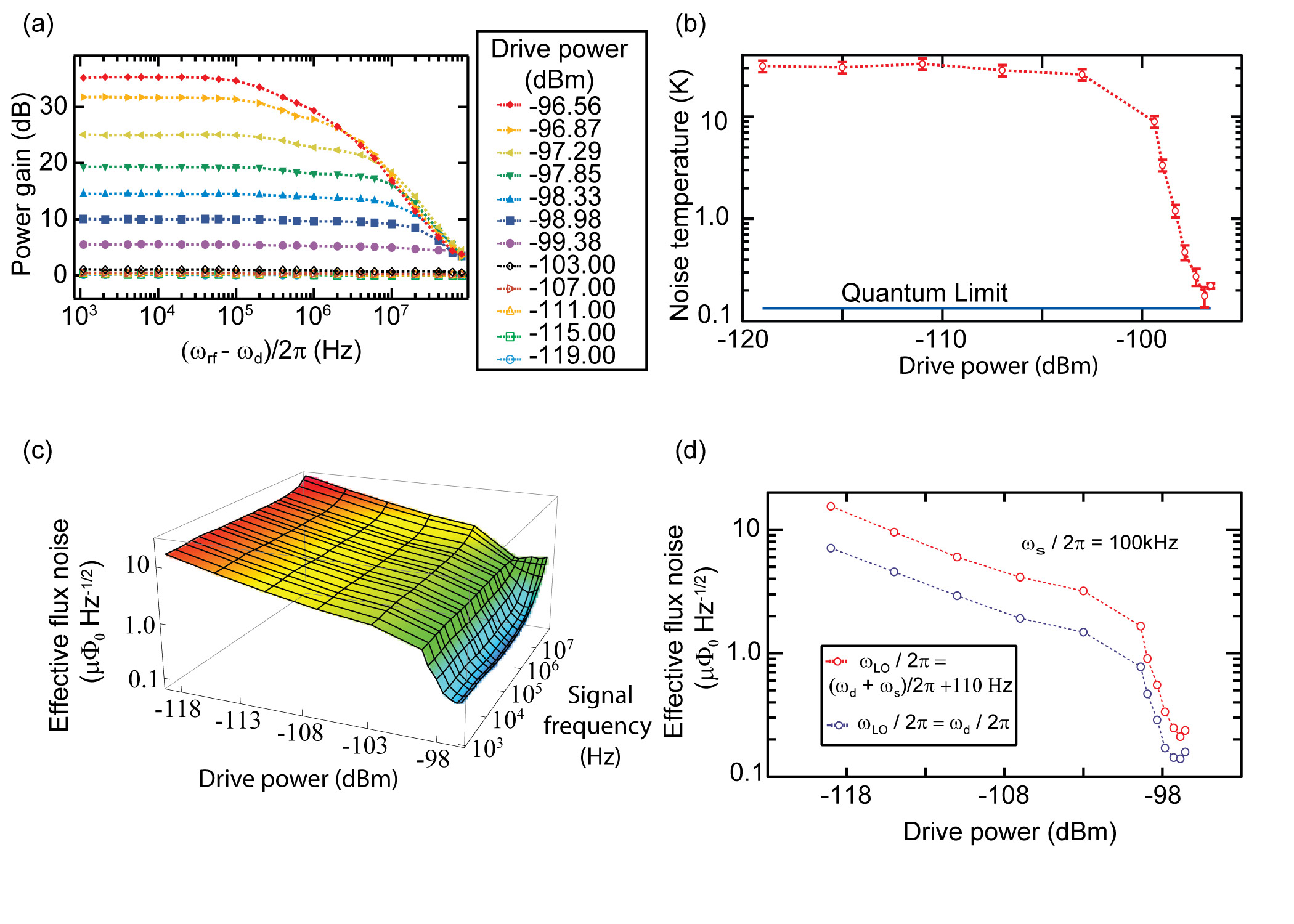}
\caption{
 (a) The parametric gain of rf signals ($\omega_{rf}/2\pi$)
applied in combination with an rf drive tone ($\omega_{d}/2\pi$) versus the offset frequency
$\left(  \omega_{rf}-\omega_{d}\right)  /2\pi$.
 \ (b) \ System noise temperature versus drive power. \ As
the parametric gain increases with drive power, the system noise temperature drops. The standard quantum limit $T_{N}=\hbar\omega _{d}/2k_{B}$ at $\omega_{d}/2\pi=5.56$\ GHz is shown as a blue horizontal line. \ (c) \ Effective flux noise versus microwave drive power and flux signal
frequency $\omega_{s}/2\pi$. As the system noise temperature decreases with drive power, the effective flux noise is reduced. \ (d) Effective flux noise versus drive power for two demodulation schemes. Data from part (c) at $\omega_{s}/2\pi$ =100 kHz and demodulated with $\omega_{LO}/2\pi=(\omega_{d}+\omega_{s})/2\pi+110$ Hz, shown in red, are compared with
those demodulated with $\omega_{LO}=\omega_{d}$, shown in blue. \ At each bias point, the phase of the LO\ signal was varied to achieve maximum sensitivity. The latter demodulation scheme shows the expected noise improvement factor of
2 for $\sqrt{G}=1$, and $\sqrt{2}$ for $\sqrt{G}>>1$.}
\end{center}
\end{figure}

\end{document}